# Factors Influencing Mode Choice of Adults with Travel-Limiting Disability[1]

**Majbah Uddin, Ph.D. (*Corresponding Author*)**
National Transportation Research Center
Oak Ridge National Laboratory
1 Bethel Valley Road, Oak Ridge, TN 37830
ORCiD: 0000-0001-9925-3881
Email: uddinm@ornl.gov

**Meiyu (Melrose) Pan, Ph.D.**
National Transportation Research Center
Oak Ridge National Laboratory
1 Bethel Valley Road, Oak Ridge, TN 37830
Email: panm@ornl.gov

**Ho-Ling Hwang, Ph.D.**
National Transportation Research Center
Oak Ridge National Laboratory
1 Bethel Valley Road, Oak Ridge, TN 37830
Email: hwanghlc@gmail.com

---

[1] This manuscript has been authored by UT-Battelle, LLC, under contract DE-AC05-00OR22725 with the US Department of Energy (DOE). The US government retains and the publisher, by accepting the article for publication, acknowledges that the US government retains a nonexclusive, paid-up, irrevocable, worldwide license to publish or reproduce the published form of this manuscript, or allow others to do so, for US government purposes. DOE will provide public access to these results of federally sponsored research in accordance with the DOE Public Access Plan (http://energy.gov/downloads/doe-public-access-plan).

**Abstract**

*Introduction:* Despite the plethora of research devoted to analyzing the impact of disability on travel behavior, not enough studies have investigated the varying impact of social and environmental factors on the mode choice of people with disabilities that restrict their ability to use transportation modes efficiently. This research gap can be addressed by investigating the factors influencing the mode choice behavior of people with travel-limiting disabilities, which can inform the development of accessible and sustainable transportation systems. Additionally, such studies can provide insights into the social and economic barriers faced by this population group, which can help policymakers to promote social inclusion and equity.

*Method:* This study utilized a Random Parameters Logit model to identify the individual, trip, and environmental factors that influence mode selection among people with travel-limiting disabilities. Using the 2017 National Household Travel Survey data for New York State, which included information on respondents with travel-limiting disabilities, the analysis focused on a sample of 8,016 people. In addition, climate data from the National Oceanic and Atmospheric Administration were integrated as additional explanatory variables in the modeling process.

*Results:* The results revealed that people with disabilities may be inclined to travel longer distances walking in the absence of suitable accommodation facilities for other transportation modes. Furthermore, people were less inclined to walk during summer and winter, indicating a need to consider weather conditions as a significant determinant of mode choice. Moreover, low-income people with disabilities were more likely to rely on public transport or walking.

*Conclusion:* Based on this study's findings, transportation agencies could design infrastructure and plan for future expansions that is more inclusive and accessible, thus catering to the mobility needs of people with travel-limiting disabilities.

1. **Introduction**

The issue of mobility challenges faced by people with disabilities is a pressing concern that warrants scholarly attention. Recent estimates suggest that a significant proportion of the American population, approximately 25.5 million people, encounter difficulties in traveling outside their homes due to their disabilities (Brumbaugh, 2022). Notably, a substantial portion of this group comprises adults aged between 18 and 64 years. Transportation serves as a fundamental element of daily life, providing access to critical services such as education, employment, healthcare, and social engagement. However, for persons with disabilities that limit their travel, identifying suitable transportation options that meet their specific needs can be a challenging task. Addressing the mobility challenges faced by people with disabilities is paramount for enhancing their quality of life, promoting independence, and advancing social equity and sustainability. Therefore, further research and policy initiatives are necessary to improve the accessibility and adequacy of transportation services for this vulnerable population group.

While much research has been done on the impact of disability on travel behavior, there is room for further exploration of how people with disabilities make decisions about transportation modes. By identifying the factors that influence their decision-making, we could enhance the mobility of people with disabilities. However, it's important to note that the transportation needs and preferences of people with disabilities are often diverse and complex (Park et al., 2023). For instance, some people may require specialized transportation services that accommodate their mobility devices or physical impairments, while others may prefer more independent modes of transportation such as private vehicles or ride-sharing services. Additionally, socioeconomic factors such as income, education level, and housing location can also significantly influence the transportation choices of adults with disabilities (Dillaway et al., 2022). Limited mobility can have a negative impact on community engagement, particularly for people with disabilities. For example, research has shown that people with blindness or low vision, psychiatric disabilities, chronic health conditions, or multiple disabilities experienced more problems using public transportation for community participation (Bezyak et al., 2020). This can further exacerbate their difficulties in participating in public engagement and communicating their mobility barriers. Despite the critical role of transportation in daily life, there is a shortage of studies that specifically address the transportation mode choice of people with disabilities. Therefore, there is an urgent need to investigate the various factors that influence the mode choice of people with travel-limiting disabilities (TLD).

Many challenges exist in investigating the mode choice influencing factors on people with disabilities (McDaniels et al., 2018). The collection of data regarding people with disabilities can be difficult, as many studies rely on focus groups that limit the number of participants and types of data collected (Lindsay & Yantzi, 2014; Ward, 2023). Additionally, trip and environment characteristics such as travel time of day, trip purpose, and weather have not been thoroughly examined. One study did analyze both survey and registration data of paratransit users and found that inclement weather conditions led to a higher usage of paratransit compared to public transit. However, the study only collected average weather data based on the travel day and region rather than users' specific departure and/or arrival times as well as origin locations (Durand & Zijlstra, 2023). This omission cannot sufficiently provide insight into the decision-making process involved in actual mode choice before a trip.

To tackle the aforementioned challenges, this study aims to investigate the multifaceted factors that affect the mode choice of people with TLD. This study utilizes the 2017 National Household Travel Survey (NHTS). Within the NHTS, a specific question, "Do you have a condition or handicap that makes it difficult to travel outside of the home?" is employed to select the target audience. Those who have answered "yes" to this question are defined as people with TLD. Accordingly, this study utilized all the trips associated with this group of respondents for analysis. The study estimated a model for identifying the individual, trip, and environmental factors that influence the mode selection of people with disabilities. The model was applied in the context of New York State. This research aims to offer a comprehensive insight into the decision-making process of people with disabilities regarding their transportation mode choice. With that, transportation agencies could better design infrastructure and plan for future expansions that is inclusive and accessible.

2. **Literature Review**

Table 1 provides a summary of the influencing factors on mode choices of people with disability in general. The mode choice of people is often influenced by health indicators, which may be related to mental or physical health concerns such as stress, mobility limitations, disability, and obesity (Mattisson et al., 2018). These factors can significantly impact travel patterns and mode choice, with people with disabilities, for example, having a lower share of non-work trips (Jansuwan et al., 2013). The influence of socioeconomic factors on mode choice has also been studied. For instance, research conducted in a developing country found that women with mobility challenges prioritize safety and travel time, even though it may result in higher transportation costs (Mogaji et al., 2023). In addition, vehicle ownership and accessibility to public transit have been shown to significantly impact mode choice (Haustein, 2012). Supporting instruments, such as walking frames, canes, crutches, and wheelchairs, also play a role in mode choice (Bhuiya et al., 2022). For example, people who use wheelchairs are more reluctant to travel by bus than those who use crutches or canes (Frye, 2013).

Various studies have explored the mode choice behavior of older people with TLD in transportation. It has been observed that older people generally undertake fewer and shorter trips compared to younger people and rely more heavily on private vehicles for their transportation needs (Khan et al., 2021a; van den Berg et al., 2011). Furthermore, studies have revealed that the interaction between age and disability also plays a significant role in the mode choice behavior of older people. Specifically, older people are more inclined to select paratransit over public transit options (Khan et al., 2021b; Schmöcker et al., 2008). Notably, mode choice behavior among older people appears to be dynamic and influenced by various factors that change over time. For instance, research indicated that older women were highly dependent on public transit, particularly when they did not have access to a personal vehicle or a transit card (Schwanen et al., 2001). However, another study found that there had been an increase in the use of personal vehicles by women in the older age group (Schwanen & Páez, 2010).

Contextual factors have been investigated in relation to mobility for people with disabilities. In winter, youth with physical disabilities face challenges participating in social and recreational activities due to limited visibility, difficulties using medical devices, and unexpected wheelchair breakdowns (Lindsay & Yantzi, 2014). Older and disabled travelers may have greater difficulty coping with adverse weather conditions than their younger counterparts (C. Liu et al., 2017), which can impede access to essential needs like food (Schwartz et al., 2023).

**Table 1. Influencing factors of mode choice of movement-challenged people.**

| Author, Date | Significant influencing factors | Group of people | Transportation Modes |
|---|---|---|---|
| Schwanen et al. 2001 | Gender, vehicle ownership | Older people | Public transport, personal vehicle |
| Schmöcker et al. 2008 | Age | Disability | Public transport (buses and trams) |
| van den Berg et al. 2011 | Trip purpose, urban density, distance, gender, education | Older people | Personal vehicle, active transportation |
| Haustein 2012 | Public transport attitudes, aspects of centrality, car availability | Disability, Older people | Personal vehicle, public transit (buses, trams, rail) |
| Jansuwan et al. 2013 | Trip characteristics, social strength, public transit accessibility | Disability | Public transport (buses), personal vehicle |
| Khan et al., 2021a | Trip purpose, departure time, distance | Older people | Personal vehicle |
| Maisel et al. 2021 | City size, built environment, bus schedules | Blind and/or visually impaired, intellectual and/or cognitive disability | Public transit |
| Khan et al., 2021b | Gender, age, vehicle ownership, household size | Disability | Paratransit |
| Bhuiya et al. 2022 | Age, sex, income, travel time, medical device | Disability | Personal vehicle, bus, walking |
| Mogaji et al. 2023 | Trip purpose, financial ability, skills for independency, security concerns | Disability | Active transportation, shared transportation |

Studies have also explored the factors influencing transportation mode choices for people with different types of disabilities. For example, people with mobility impairments prioritize built environment factors over scheduling-related factors when deciding on transit modes, while riders with intellectual and cognitive disabilities require assistance with complex trips (Maisel et al., 2021). However, it is worth considering that bicycles can provide certain advantages to individuals with disabilities which do not have a large impact on their movement. According to a focus group study, people with hearing disabilities were more inclined to use bicycles than public transit, as bicycle give them a higher level of autonomy (Mogaji et al., 2023). However, not many studies have focused on the transportation mode choices of people with TLD, whose decision-making processes may differ from those with other types of disabilities.

2.1 Summary of Gaps and Contributions of this Research

In essence, the existing research on the mode preferences of people with TLD lacks sufficient investigation into the influence of weather conditions. While the mode choices of TLD people have been extensively explored, certain aspects, like weather impacts, remain underexplored. Although qualitative analyses have suggested a heightened sensitivity of TLD people's mode choices to weather conditions (Lindsay & Yantzi, 2014), quantitative methods to measure this effect are lacking. While specific modes, such as accessible taxis, have been studied in relation to weather condition (Zhang et al., 2023), there's a need for targeted research encompassing multimodal transportation.

To this end, this study aims to assess how weather conditions at trip start and end times and locations impact the mode choices of TLD people, encompassing options such as personal vehicles, public transportation, and paratransit. This research offers two main contributions. Firstly, it employs the NHTS dataset, ensuring reproducibility across different regions. Leveraging this extensive survey data allows for the analysis of a large number of trips, providing an advantage over smaller-scale and relatively costly focus groups or survey studies. Secondly, the study combines historical weather data with trip data to capture the actual decision-making context of TLD people. By employing a random parameters logit model, the study has the capability to quantify the impact of weather conditions. This holistic approach considers how weather conditions at both the commencement and conclusion of trips shape the process of mode selection for TLD people.

## 3. Materials and Methods

3.1 Data Description

*3.1.1 National Household Travel Survey (NHTS)*

The 2017 NHTS is used as the primary data source. The survey is conducted by the Federal Highway Administration, US Department of Transportation, and is the authoritative source on the travel behavior of the American public (Federal Highway Administration, 2017). It also has a robust history of use in transportation research, particularly in understanding travel behavior and patterns. In this study, surveyed households that are located in New York State were used. The information was gathered by the NHTS for a total of 17,209 households, 35,967 persons, and 120,414 trips for the state of New York.

Four modes of transportation, namely Personal Vehicle, Walk, Public Transport, and Other Mode were considered in this study. Public Transport was defined to include public bus, city-to-city bus, Amtrak, and subway. Other Mode mainly referred to other transportation services, such as paratransit, taxi, and private bus. The share of the modes of transportation is shown in Table 2.

**Table 2. Share of modes of transportation included in "Other Mode" category**

| NHTS mode | Proportion |
|---|---|
| Taxi | 29.2% |
| Something else | 21.5% |
| Paratransit | 18.2% |
| Bicycle | 6.2% |
| School bus | 6.2% |
| Private bus | 5.5% |
| Rental car | 4.4% |
| Airplane | 3.3% |
| Golf cart | 2.6% |
| Recreational vehicle (RV) | 2.2% |
| Boat | 0.7% |

The study utilized two types of data from the NHTS as explanatory variables. The first category encompasses demographic features, comprising age, gender, race, ethnicity, working status, income, health condition, medical devices, and education. The second category entails trip attributes, including whether the individual is driving, the purpose of the trip, day of the week, urban/rural, season, loop trip, and whether origins and destinations are in New York City vs rest of New York State. These variables were selected due to their potential impact on mode choice (Jansuwan et al., 2013; Park et al., 2023) as well as better coverage in the data set. For example, studies have indicated that residents of New York City are more inclined to walking compared to their counterparts from other regions of the state (Y. Liu et al., 2022), and non-worker disabled people are less likely to use public transit than others (Kwon & Akar, 2022). Altogether 27 explanatory variables from NHTS were explored.

*3.1.2 Climate Data Online*

Weather data were collected by extracting information from the National Climate Data Center (NCDC) available through Climate Data Online (CDO) (NOAA CDO Climate Data Online (CDO), 2023). The NCDC archives weather data from various sources, including radar, satellites, airport and military weather stations across the nation. The CDO station-level hourly weather data were utilized to identify nearby weather conditions at the trip start and end location based on trip start and end time. The absolute difference in time between trip start time and weather station time was used to determine the weather at trip start and end times. Five variables were utilized as explanatory variables, i.e., temperature, precipitation, humidity, visibility, and wind speed, which were then merged with the NHTS based on trip origin and start time and trip destination and end time.

3.2 Definition of Disability

In the 2017 NHTS, a person with TLD was defined as one who answers "yes" to the questions of: *Do you have a condition or handicap that makes it difficult to travel outside of the home?* Figure 1 (a) presents the compensating mobility strategies of people with TLD. According to the NHTS, over 65% of survey respondents with TLD reported that they reduced their day-to-day travel. However, the survey did not further investigate the types of trips or circumstances under which people gave up traveling. Therefore, this study focuses on the decision-making processes of those with TLD when they do travel. The medical devices used by people with TLD are primarily walking canes and walkers, as indicated in Figure 1 (b).

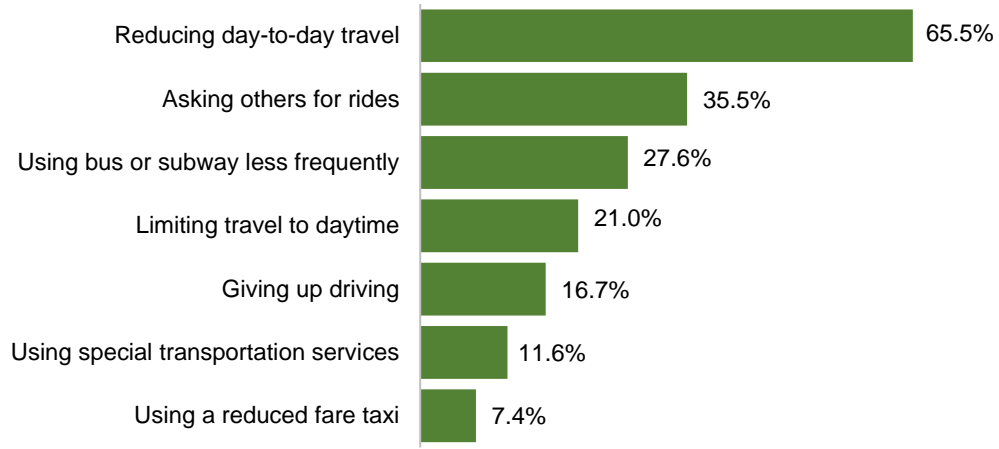

(a) Share of compensating strategies.

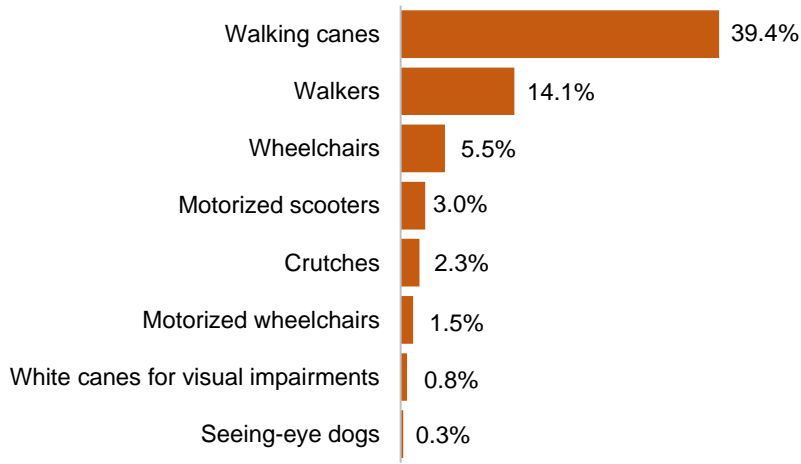

(b) Share of medical devices.

**Figure 1. Share of compensating strategies and medical devices of NHTS respondents with TLD.**

3.3 Random Parameters Logit Model

Given that people with TLD may encounter diverse obstacles due to their unique health conditions and situational contexts, their mode choice may vary significantly between people, making fixed parameter models inappropriate. As a result, this study utilized Random Parameters Logit (RPL) modeling, as its use is necessary to account for unobserved heterogeneity.

The present study seeks to examine the relationship between travel mode choices made by people with TLD and relevant explanatory variables. The relationship can be expressed through the following equation:

$$Y_{ij} = \boldsymbol{\beta}_i \boldsymbol{X}_{ij} + \varepsilon_{ij}$$

where $Y_{ij}$ denotes individual $j$'s mode choice ($i \in I$ where $I$={Personal Vehicle, Public Transport, Walk, Other Mode}), $X_{ij}$ represents the value of the independent variable $X$ for individual $j$ for mode choice $i$, $\boldsymbol{\beta}_i$ represents a vector of estimable parameter for mode choice $i$. Assuming the error term is independently and identically distributed with a generalized extreme value distribution, the resulting model conforms to a standard multinomial logit model. The choice probability $P_j(i)$ of individual $j$ choosing mode $i$ is given by the following equation:

$$P_j(i) = \frac{\exp(\beta_i X_{ij})}{\sum_{i \in I} \exp(\beta_i X_{ij})}$$

The probability of choosing mode $i$ is determined by integrating the conditional probability over all possible values of $\beta_i$, which represents the preference of an individual for that mode. The resulting choice probability is a weighted average of the standard multinomial logit probabilities:

$$P_j(i|\theta) = \int \frac{\exp(\beta_i X_{ij})}{\sum_{i \in I} \exp(\beta_i X_{ij})} f(\beta_i|\theta) d\beta_i$$

where $P_j(i|\theta)$ denotes the probability of choosing mode $i$ conditional on $f(\beta_i|\theta)$, where $f(\beta_i|\theta)$ represents the density function of $\beta_i$ and $\theta$ is a vector of parameters to be estimated of this density function. The density function $f(\beta_i|\theta)$, which represents the distribution of individual preferences for a given mode of transportation, can take any form. In this study, the normal distribution is employed as it facilitates a better interpretation of the results (Milton et al., 2008).

To estimate the parameters of the density function $f(\beta_i|\theta)$, which describes the distribution of individual preferences, a maximum likelihood estimation is performed using a simulation-based approach. To enhance the efficiency of the numerical integration process, Halton draws are utilized. Prior research has demonstrated that Halton draws are more efficient and require fewer draws to achieve convergence compared to other methods such as random draws (Bhat, 2003; Train, 2009). Our reported results are based on 200 Halton draws.

Marginal effects were also calculated in this study to provide additional information about the impact of explanatory variables on the probability of choosing a specific mode of transportation. While model coefficients inform the direction and magnitude of the relationship between the explanatory variables and mode choice, marginal effects measure the change in probability associated with a unit change in an explanatory variable, holding other variables constant.

## 4. Results

4.1 Data Description

Out of all the people who participated in the survey, those with TLD and an age of 18 years or older were chosen for inclusion. The final dataset comprised a total of 8,016 people. Table 3 shows the descriptive statistics of the binary variables, including sample size and the percentage of each category. Table 4 shows the descriptive statistics of the continuous variables, including sample size, unit, mean, and standard deviation. The dependent variable is the mode choice of a person with TLD.

**Table 3. Descriptive statistics of binary variables.**

| Variable | Personal Vehicle | | Public Transport | | Walk | | Other Mode | | Total | |
|---|---|---|---|---|---|---|---|---|---|---|
| Total | 6,491 | 81.0% | 257 | 3.2% | 994 | 12.4% | 274 | 3.4% | 8,016 | 100.0% |
| *Age* | | | | | | | | | | |
| 18-24 | 77 | 65.3% | 4 | 3.4% | 17 | 14.4% | 20 | 16.9% | 118 | 1.5% |
| 25-44 | 444 | 69.4% | 37 | 5.8% | 123 | 19.2% | 36 | 5.6% | 640 | 8.0% |
| 45-64 | 2,406 | 78.7% | 143 | 4.7% | 407 | 13.3% | 103 | 3.4% | 3,059 | 38.2% |
| Over 65 | 3,564 | 84.9% | 73 | 1.7% | 447 | 10.6% | 115 | 2.7% | 4,199 | 52.4% |
| *Gender* | | | | | | | | | | |
| Male | 2,810 | 82.3% | 111 | 3.3% | 395 | 11.6% | 97 | 2.8% | 3,413 | 42.6% |
| Female | 3,681 | 80.0% | 146 | 3.2% | 599 | 13.0% | 177 | 3.8% | 4,603 | 57.4% |
| *Worker* | | | | | | | | | | |
| Yes | 1,077 | 81.9% | 49 | 3.7% | 159 | 12.1% | 30 | 2.3% | 1,315 | 16.4% |
| No | 5,414 | 80.8% | 208 | 3.1% | 835 | 12.5% | 244 | 3.6% | 6,701 | 83.6% |
| *Driver* | | | | | | | | | | |
| Yes | 4,279 | 99.7% | 0 | 0.0% | 0 | 0.0% | 12 | 0.3% | 4,291 | 53.5% |
| No | 2,212 | 59.4% | 257 | 6.9% | 994 | 26.7% | 262 | 7.0% | 3,725 | 46.5% |
| *Race* | | | | | | | | | | |
| White | 5,820 | 84.1% | 143 | 2.1% | 747 | 10.8% | 211 | 3.0% | 6,921 | 90.1% |
| Non-white | 432 | 56.7% | 100 | 13.1% | 178 | 23.4% | 52 | 6.8% | 762 | 9.9% |
| *Hispanic/Latino* | | | | | | | | | | |
| Yes | 203 | 56.1% | 44 | 12.2% | 88 | 24.3% | 27 | 7.5% | 362 | 4.5% |
| No | 6,246 | 82.2% | 211 | 2.8% | 899 | 11.8% | 247 | 3.2% | 7,603 | 95.5% |

| Variable | Personal Vehicle | | Public Transport | | Walk | | Other Mode | | Total | |
|---|---|---|---|---|---|---|---|---|---|---|
| *Educational attainment* | | | | | | | | | | |
| Less than bachelor | 4,556 | 81.2% | 187 | 3.3% | 653 | 11.6% | 212 | 3.8% | 5,608 | 70.0% |
| Bachelor or higher | 1,931 | 80.5% | 70 | 2.9% | 336 | 14.0% | 62 | 2.6% | 2,399 | 30.0% |
| *Household income* | | | | | | | | | | |
| <$50,000 | 3,867 | 77.4% | 200 | 4.0% | 721 | 14.4% | 205 | 4.1% | 4,993 | 64.4% |
| $50,000 to $74,999 | 1,072 | 88.5% | 24 | 2.0% | 89 | 7.3% | 26 | 2.1% | 1,211 | 15.6% |
| $75,000 to $99,999 | 535 | 91.0% | 3 | 0.5% | 42 | 7.1% | 8 | 1.4% | 588 | 7.6% |
| $100,000 to $199,999 | 682 | 84.6% | 15 | 1.9% | 86 | 10.7% | 23 | 2.9% | 806 | 10.4% |
| $200,000 or more | 124 | 81.6% | 3 | 2.0% | 21 | 13.8% | 4 | 2.6% | 152 | 2.0% |
| *Trip origin location* | | | | | | | | | | |
| Rural | 3,400 | 87.7% | 33 | 0.9% | 368 | 9.5% | 74 | 1.9% | 3,875 | 48.6% |
| Urban | 3,047 | 74.5% | 224 | 5.5% | 621 | 15.2% | 200 | 4.9% | 4,092 | 51.4% |
| *Trip destination location* | | | | | | | | | | |
| Rural | 3,392 | 87.6% | 34 | 0.9% | 369 | 9.5% | 78 | 2.0% | 3,873 | 48.6% |
| Urban | 3,058 | 74.6% | 223 | 5.4% | 620 | 15.1% | 196 | 4.8% | 4,097 | 51.4% |
| *Day of week* | | | | | | | | | | |
| Weekday | 4,969 | 80.5% | 204 | 3.3% | 767 | 12.4% | 234 | 3.8% | 6,174 | 77.0% |
| Weekend | 1,522 | 82.6% | 53 | 2.9% | 227 | 12.3% | 40 | 2.2% | 1,842 | 23.0% |
| *Trip purpose* | | | | | | | | | | |
| Work | 306 | 80.5% | 25 | 6.6% | 30 | 7.9% | 19 | 5.0% | 380 | 4.7% |
| Non-work | 6,184 | 81.0% | 231 | 3.0% | 964 | 12.6% | 255 | 3.3% | 7,634 | 95.3% |
| *Loop trip* | | | | | | | | | | |
| Yes | 19 | 8.5% | 0 | 0.0% | 201 | 89.7% | 4 | 1.8% | 224 | 2.8% |
| No | 6,472 | 83.1% | 257 | 3.3% | 793 | 10.2% | 270 | 3.5% | 7,792 | 97.2% |
| *Trip category* | | | | | | | | | | |
| Home-based | 4,147 | 78.3% | 194 | 3.7% | 752 | 14.2% | 200 | 3.8% | 5,293 | 66.0% |
| Non-home-based | 2,344 | 86.1% | 63 | 2.3% | 242 | 8.9% | 74 | 2.7% | 2,723 | 34.0% |
| *Born in the U.S.* | | | | | | | | | | |
| Yes | 6,129 | 82.3% | 210 | 2.8% | 860 | 11.5% | 248 | 3.3% | 7,447 | 92.9% |
| No | 362 | 63.6% | 47 | 8.3% | 134 | 23.6% | 26 | 4.6% | 569 | 7.1% |
| *Health condition* | | | | | | | | | | |
| Poor health | 618 | 77.2% | 21 | 2.6% | 115 | 14.4% | 47 | 5.9% | 801 | 10.0% |
| Not poor health | 5,873 | 81.4% | 236 | 3.3% | 879 | 12.2% | 227 | 3.1% | 7,215 | 90.0% |
| *Time of day* | | | | | | | | | | |
| 7:00 am to 9:59 am | 1,006 | 79.0% | 50 | 3.9% | 148 | 11.6% | 70 | 5.5% | 1,274 | 15.9% |
| 10:00 am to 3:59 pm | 3,762 | 81.9% | 148 | 3.2% | 542 | 11.8% | 140 | 3.0% | 4,592 | 57.3% |
| 4:00 pm to 6:59 pm | 1,118 | 81.5% | 34 | 2.5% | 182 | 13.3% | 37 | 2.7% | 1,371 | 17.1% |
| 7:00 pm to 6:59 am | 605 | 77.7% | 25 | 3.2% | 122 | 15.7% | 27 | 3.5% | 779 | 9.7% |
| *Season* | | | | | | | | | | |

| Variable | Personal Vehicle | | Public Transport | | Walk | | Other Mode | | Total | |
|---|---|---|---|---|---|---|---|---|---|---|
| Summer | 1,646 | 80.6% | 65 | 3.2% | 262 | 12.8% | 68 | 3.3% | 2,041 | 25.5% |
| Fall | 1,715 | 79.9% | 76 | 3.5% | 285 | 13.3% | 71 | 3.3% | 2,147 | 26.8% |
| Winter | 1,897 | 83.2% | 75 | 3.3% | 233 | 10.2% | 74 | 3.2% | 2,279 | 28.4% |
| Spring | 1,233 | 79.6% | 41 | 2.6% | 214 | 13.8% | 61 | 3.9% | 1,549 | 19.3% |
| *Trip origin in New York City (NYC)* | | | | | | | | | | |
| Yes | 289 | 44.9% | 109 | 17.0% | 199 | 30.9% | 46 | 7.2% | 643 | 8.0% |
| No | 6,202 | 84.1% | 148 | 2.0% | 795 | 10.8% | 228 | 3.1% | 7,373 | 92.0% |
| *Trip destination in NYC* | | | | | | | | | | |
| Yes | 312 | 44.6% | 116 | 16.6% | 221 | 31.6% | 51 | 7.3% | 700 | 8.7% |
| No | 6,179 | 84.5% | 141 | 1.9% | 773 | 10.6% | 223 | 3.0% | 7,316 | 91.3% |
| *Working from home* | | | | | | | | | | |
| Yes | 196 | 89.9% | 2 | 0.9% | 15 | 6.9% | 5 | 2.3% | 218 | 2.7% |
| No | 6,295 | 80.7% | 255 | 3.3% | 979 | 12.6% | 269 | 3.4% | 7,798 | 97.3% |
| *Medical devices* | | | | | | | | | | |
| *Cane* | | | | | | | | | | |
| Yes | 2,748 | 81.3% | 125 | 3.7% | 393 | 11.6% | 114 | 3.4% | 3,380 | 42.2% |
| No | 3,743 | 80.7% | 132 | 2.8% | 601 | 13.0% | 160 | 3.5% | 4,636 | 57.8% |
| *Manual wheelchair* | | | | | | | | | | |
| Yes | 419 | 78.6% | 25 | 4.7% | 29 | 5.4% | 60 | 11.3% | 533 | 6.6% |
| No | 6,072 | 81.1% | 232 | 3.1% | 965 | 12.9% | 214 | 2.9% | 7,483 | 93.4% |
| *Crutch* | | | | | | | | | | |
| Yes | 211 | 84.7% | 9 | 3.6% | 20 | 8.0% | 9 | 3.6% | 249 | 3.1% |
| No | 6,280 | 80.9% | 248 | 3.2% | 974 | 12.5% | 265 | 3.4% | 7,767 | 96.9% |
| *Dog assistance* | | | | | | | | | | |
| Yes | 28 | 50.9% | 9 | 16.4% | 11 | 20.0% | 7 | 12.7% | 55 | 0.7% |
| No | 6,463 | 81.2% | 248 | 3.1% | 983 | 12.3% | 267 | 3.4% | 7,961 | 99.3% |

Note: The first column under each category presents the sample size and the second column presents the percentage of each category.

**Table 4. Descriptive statistics of continuous variables.**

| Variable | Statistics | Unit | Personal Vehicle | Public Transport | Walk | Other Mode | Average |
|---|---|---|---|---|---|---|---|
| Log(Trip length) | Mean | mile | 1.14 | 1.23 | −1.29 | 1.2 | 0.85 |
| | Standard deviation | mile | 1.34 | 1.14 | 1.23 | 1.8 | 1.56 |
| | Sample size | | 6,489 | 255 | 988 | 272 | 8,004 |
| Log(Trip duration) | Mean | minute | 2.58 | 3.73 | 2.46 | 3.19 | 2.63 |
| | Standard deviation | minute | 0.85 | 0.75 | 1.02 | 0.95 | 0.91 |
| | Sample size | | 6,480 | 257 | 993 | 272 | 8,002 |
| *Specific to trip origin* | | | | | | | |
| Temperature | Mean | °F | 55.0 | 58.4 | 55.9 | 55.1 | 55.2 |
| | Standard deviation | °F | 20.5 | 19.8 | 19.9 | 19.4 | 20.4 |
| | Sample size | | 6,458 | 257 | 987 | 274 | 7,976 |
| Precipitation | Mean | inch | 0.3 | 0.2 | 0.4 | 0.2 | 0.3 |
| | Standard deviation | inch | 2.0 | 1.0 | 1.9 | 1.0 | 1.9 |
| | Sample size | | 4,928 | 206 | 734 | 198 | 6,066 |
| Humidity | Mean | % | 62.9 | 59.8 | 62.4 | 62.8 | 62.7 |
| | Standard deviation | % | 18.7 | 17.9 | 19.7 | 18 | 18.8 |
| | Sample size | | 6,455 | 257 | 987 | 274 | 7,973 |
| Visibility | Mean | mile | 9.4 | 9.5 | 9.2 | 9.3 | 9.4 |

| Variable | Statistics | Unit | Personal Vehicle | Public Transport | Walk | Other Mode | Average |
|---|---|---|---|---|---|---|---|
| | Standard deviation | mile | 2.6 | 2.7 | 2.8 | 2.5 | 2.7 |
| | Sample size | | 6,442 | 255 | 987 | 272 | 7,956 |
| Wind speed | Mean | mph | 8.9 | 9.1 | 8.6 | 8.7 | 8.8 |
| | Standard deviation | mph | 5.6 | 5.7 | 5.7 | 5.9 | 5.6 |
| | Sample size | | 6,406 | 241 | 978 | 274 | 7,899 |
| *Specific to trip destination* | | | | | | | |
| Temperature | Mean | °F | 55 | 58.5 | 56 | 55.2 | 55.3 |
| | Standard deviation | °F | 20.5 | 20 | 19.9 | 19.3 | 20.4 |
| | Sample size | | 6,461 | 257 | 988 | 274 | 7,980 |
| Precipitation | Mean | inch | 0.3 | 0.1 | 0.4 | 0.4 | 0.3 |
| | Standard deviation | inch | 1.9 | 0.5 | 1.9 | 1.7 | 1.9 |
| | Sample size | | 4,890 | 206 | 733 | 207 | 6,036 |
| Humidity | Mean | % | 62.7 | 59.1 | 62.1 | 62.7 | 62.5 |
| | Standard deviation | % | 18.7 | 17.9 | 19.5 | 18 | 18.8 |
| | Sample size | | 6,458 | 257 | 988 | 274 | 7,977 |
| Visibility | Mean | mile | 9.4 | 9.4 | 9.3 | 9.3 | 9.4 |
| | Standard deviation | mile | 2.6 | 2.6 | 2.7 | 2.4 | 2.6 |
| | Sample size | | 6,447 | 255 | 988 | 272 | 7,962 |
| Wind speed | Mean | mph | 8.8 | 9.1 | 8.7 | 8.9 | 8.8 |
| | Standard deviation | mph | 5.6 | 5.5 | 5.7 | 5.9 | 5.6 |
| | Sample size | | 6,408 | 243 | 975 | 272 | 7,898 |

### 4.2 Model Results

During the model development process, variables were retained in the specification if they have *t*-statistics corresponding to the 95% confidence level or higher on a two-tailed *t*-test. The random parameters were retained if their standard deviations have *t*-statistics corresponding to the 95% confidence level or higher. A summary of the coefficients of the significant variables and RPL model results are shown in Table 5. The pseudo R-squared from the model outcome is 0.74, indicating a good model fit.

      A positive coefficient value for an explanatory variable indicates a positive association with the mode choice and increases the probability of selecting that particular mode. For instance, higher income people with TLD was positively associated with a higher probability of choosing personal vehicles, while lower income were more likely to select public transport or walking as their preferred mode of travel. It is also found that public transport is more likely to be chosen when wind speed is higher.

Table 5. Summary of RPL model results.

| Variable | Coefficient | t-statistics | p-value |
|---|---|---|---|
| *Defined for Personal Vehicle* | | | |
| Constant | 8.33 | 19.72 | 0.000 |
| Age: Over 65 | 0.90 | 9.43 | 0.000 |
| Race: White | 0.84 | 7.11 | 0.000 |
| Household Income: $100,000 to $199,999 | 0.54 | 2.46 | 0.014 |
| Day of week: Weekday | −0.26 | −2.36 | 0.018 |
| Trip category: Home-based | −1.20 | −9.61 | 0.000 |
| Trip Destination in NYC | −1.16 | −8.00 | 0.000 |
| Medical device: Walking cane | 0.43 | 4.10 | 0.000 |
| Log(Trip duration) | −2.10 | −27.00 | 0.000 |
| Log(Trip length) | 0.81 | 14.16 | 0.000 |
| *Defined for Public Transport* | | | |
| Constant | −2.73 | −6.29 | 0.000 |
| Household Income: <$50,000 | 0.90 | 4.76 | 0.000 |
| Trip destination location: Urban | 1.08 | 5.07 | 0.000 |
| Trip purpose: Work | 0.75 | 2.65 | 0.008 |
| Trip category: Non-home-based | 0.74 | 3.69 | 0.000 |
| Season: Fall | 0.36 | 2.21 | 0.027 |
| Trip origin in NYC | 0.88 | 4.51 | 0.000 |
| Medical device: Walking cane | 0.61 | 3.58 | 0.000 |
| Trip origin: Wind speed | 0.03 | 2.19 | 0.029 |
| *Defined for Walk* | | | |
| Constant | 1.00 | 2.89 | 0.004 |
| Hispanic/Latino | 0.51 | 2.42 | 0.016 |
| Household Income: <$50,000 | 0.40 | 3.07 | 0.002 |
| Season: Summer | −0.31 | −2.33 | 0.020 |
| Season: Winter | −0.29 | −2.22 | 0.026 |
| Log(Trip length) | −1.58 | −24.43 | 0.000 |
| *Defined for Other Mode (e.g., paratransit, taxi)* | | | |
| Educational attainment: Bachelor or higher | −0.74 | −2.97 | 0.003 |
| Trip origin location: Rural | −1.06 | −4.23 | 0.000 |
| Trip purpose: Non-work | −2.54 (2.84) | −4.13 (6.96) | 0.000 (0.000) |
| Trip category: Non-home-based | 0.70 | 2.77 | 0.006 |
| Time of day: 7:00 am to 9:59 am | 0.91 | 3.64 | 0.000 |
| Season: Summer | −1.04 (1.95) | −1.29 (2.24) | 0.198 (0.025) |
| Medical device: Wheelchair | 2.42 | 7.13 | 0.000 |
| *Model Statistics* | | | |
| Number of observations | 7,873 | | |
| Log-likelihood at zero, $LL(0)$ | −10,914.30 | | |
| Log-likelihood at convergence, $LL(\beta)$ | −2,787.27 | | |
| $\rho^2 = 1 - LL(\beta)/LL(0)$ | 0.74 | | |

The random variable results, the mean and standard deviation of the coefficients, have a distinct interpretation compared to the model coefficients. The findings presented in Table 5 demonstrate that two variables have random effects with respect to other transportation services, such as paratransit or taxi. Figure 2 further shows the probability of people choosing other transportation services given the two variables. Specifically, the non-work trip purpose variable had a mean of -2.54 and standard deviation of 2.84. This indicated that in 81.42% of the cases where non-work trips were taken, the probability of choosing other modes was reduced. The summer season variable had a mean of -1.04 and standard deviation of 1.95. This suggested that in 70.3% of the cases where trips were made during summer, the probability of selecting other modes was reduced.

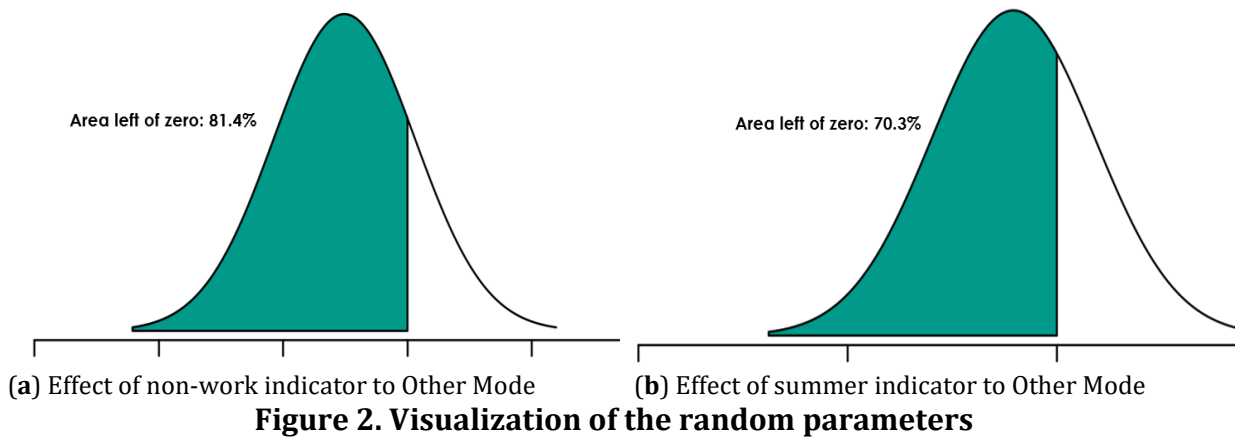

(**a**) Effect of non-work indicator to Other Mode  (**b**) Effect of summer indicator to Other Mode

**Figure 2. Visualization of the random parameters**

4.3 Marginal Effects

Table 6 displays the marginal effects of all variables included in the models. For personal vehicles, the marginal effect of race white was 0.0458, indicating that white people with TLD were 4.58% more inclined than their nonwhite counterparts to opt for personal vehicles. Moreover, an increase in the log of trip length was associated with a marginal effect of 0.0172 for personal vehicle use, which implies that a 1.72% rise in the propensity to use personal vehicles would result from an increase in the trip length. Regarding public transport, the marginal effect suggested that the propensity for its use was 2.10% higher in urban areas compared to other areas. Households with incomes lower than $50K demonstrated a 1.57% higher inclination towards using public transport than those with higher incomes.

Table 6. Marginal effects of the significant variables.

| Variable | Personal Vehicle | Public Transport | Walk | Other Mode |
|---|---|---|---|---|
| *Defined for Personal Vehicle* | | | | |
| Age: Over 65 | 0.0263 | −0.0057 | −0.0149 | −0.0057 |
| Race: White | 0.0458 | −0.0097 | −0.0257 | −0.0104 |
| Household Income: $100,000 to $199,999 | 0.0018 | −0.0004 | −0.0009 | −0.0005 |
| Day of week: Weekday | −0.0135 | 0.0032 | 0.0073 | 0.0030 |
| Trip category: Home-based | −0.0568 | 0.0140 | 0.0303 | 0.0125 |
| Trip Destination in NYC | −0.0121 | 0.0049 | 0.0055 | 0.0017 |
| Medical device: Walking cane | 0.0118 | −0.0033 | −0.0061 | −0.0024 |
| Log(Trip duration) | −0.3858 | 0.1125 | 0.1791 | 0.0942 |
| Log(Trip length) | 0.0172 | −0.0193 | 0.0191 | −0.0170 |
| *Defined for Public Transport* | | | | |
| Household Income: <$50,000 | −0.0106 | 0.0157 | −0.0039 | −0.0012 |
| Trip destination location: Urban | −0.0140 | 0.0210 | −0.0052 | −0.0018 |
| Trip purpose: Work | −0.0010 | 0.0015 | −0.0002 | −0.0003 |
| Trip category: Non-home-based | −0.0032 | 0.0043 | −0.0008 | −0.0003 |
| Season: Fall | −0.0018 | 0.0025 | −0.0006 | −0.0002 |
| Trip origin in NYC | −0.0035 | 0.0064 | −0.0021 | −0.0008 |
| Medical device: Walking cane | −0.0046 | 0.0066 | −0.0016 | −0.0005 |
| Trip origin: Wind speed | −0.0042 | 0.0061 | −0.0014 | −0.0005 |
| *Defined for Walk* | | | | |
| Hispanic/Latino | −0.0012 | −0.0004 | 0.0018 | −0.0002 |
| Household Income: <$50,000 | −0.0103 | −0.0017 | 0.0129 | −0.0009 |
| Season: Summer | 0.0029 | 0.0004 | −0.0035 | 0.0002 |
| Season: Winter | 0.0029 | 0.0004 | −0.0035 | 0.0002 |
| Log(Trip length) | −0.0372 | −0.0010 | 0.0408 | −0.0026 |
| *Defined for Other Mode (e.g., paratransit, taxi)* | | | | |
| Educational attainment: Bachelor or higher | 0.0025 | 0.0003 | 0.0005 | −0.0033 |
| Trip origin location: Rural | 0.0054 | 0.0002 | 0.0007 | −0.0063 |
| Trip purpose: Non-work | −0.0183 | −0.0002 | −0.0030 | 0.0215 |
| Trip category: Non-home-based | −0.0031 | −0.0003 | −0.0004 | 0.0038 |
| Time of day: 7:00 am to 9:59 am | −0.0032 | −0.0004 | −0.0006 | 0.0042 |
| Season: Summer | −0.0023 | −0.0001 | −0.0004 | 0.0028 |
| Medical device: Wheelchair | −0.0062 | −0.0003 | −0.0004 | 0.0038 |

## 5. Discussions

5.1 Summary of Findings

Table 7 summarizes the key research findings on the likelihood of increasing a specific mode choice and the contributing factors. People with TLD were more likely to use personal vehicles when they have a relatively higher income, are older, or are white. In contrast, lower-income people were more inclined to use walking or public transit as their mode of transportation, which was consistent with prior research on people without disabilities who had a higher rate of driving alone among higher-income and white people (Martens et al., 2019; McKenzie, 2015). This suggests potential issues regarding the affordability of various modes of transportation for people with TLD. Furthermore, prior studies indicates that people feel more autonomous when traveling using personal vehicles, which raises concerns about limited transportation mode options and lower autonomy faced by people with TLD. Moreover, this current study found that people with TLD were more likely to use public transport in urban areas, in New York City, or for occupational purposes. This predilection may be attributed to the fact that disabled passengers residing in rural regions are generally underserved by public transportation options, especially when it comes to long-distance commutes between their residences and workplaces located at a considerable distance (Watermeyer et al., 2018). This trend aligns with the denser concentration of public transit infrastructure in urban areas, as evident in Figure 3, which shows the overlay of public transit stops within urban counties in New York State (US Department of Transportation, 2023). As expected, New York City prominently shows an elevated density of public transit stops dispersed extensively across the whole city.

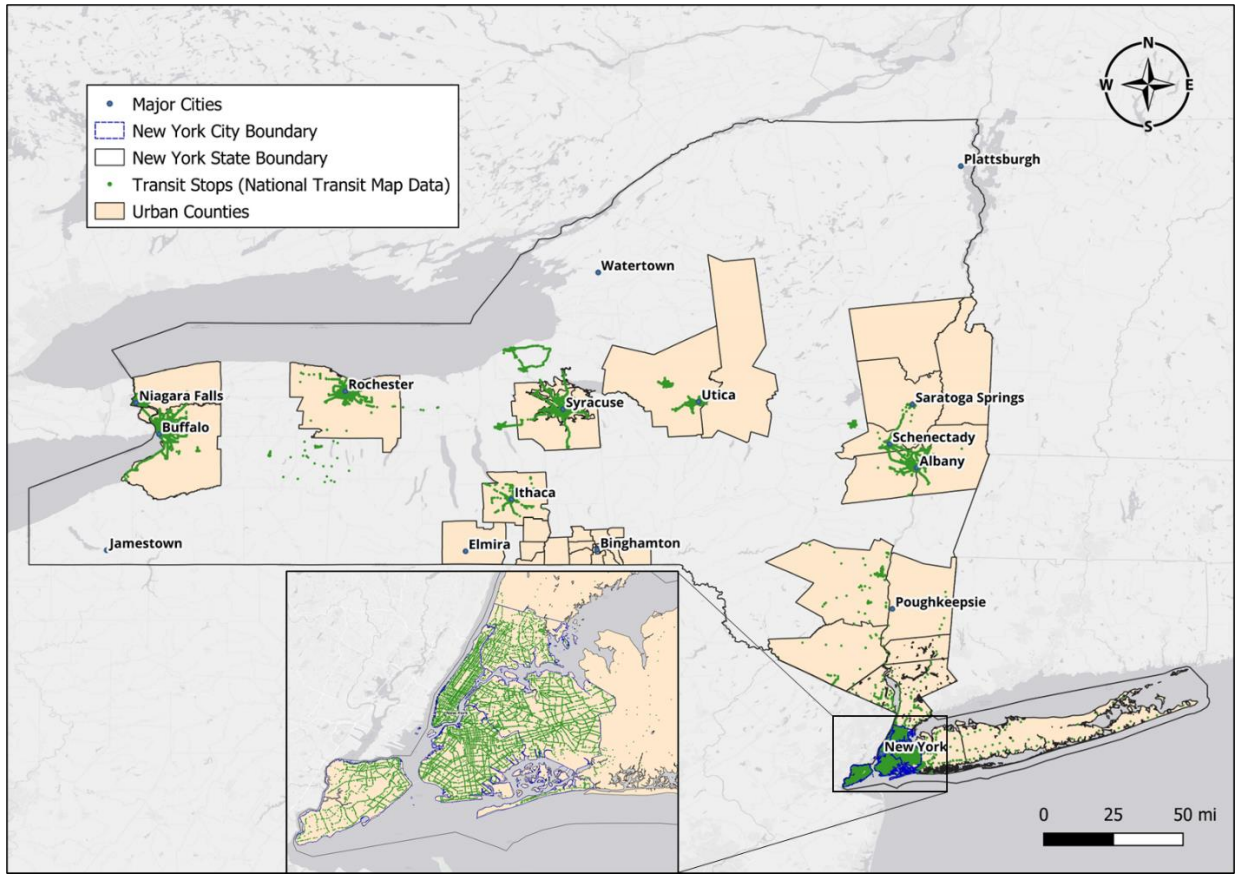

**Figure 3. Public transit stops in New York State**

The study also discovered that people were less likely to walk compared to the other three modes of transportation as the trip distance increased in general. However, the marginal effects indicated that an increase in the trip distance could lead to a slight increase in the propensity of walking. The reason for this could be a lack of offboarding and/or onboarding accommodations for their medical devices such as wheelchairs, leading them to prefer walking for longer distances, if the destination is still within a manageable walking distance. This finding is supported by previous research that people place the heaviest weight on the accessibility of accommodation facilities to maximize their travel satisfaction, while public transport is not always accessible and convenient for disabled commuters with wheelchairs (Lyu, 2017; Mogaji et al., 2023). Extreme weather conditions, such as some days in summer and winter, can also discourage people from walking. As people with lower income tend to opt for walking as a primary mode of transportation, their mobility choices are more susceptible to the impact of weather conditions, which could potentially curtail their travel options.

Lastly, people with TLD in rural areas were found to use paratransit or taxis less frequently. This could potentially be due to these services' non-operation or less accessibility in regions where the population density is relatively low (Lewis & Regan, 2020).

**Table 7. Summary of positive and negative relationship.**

| Type | Variable | Personal Vehicle | Public Transport | Walk | Other Mode |
|---|---|---|---|---|---|
| Individual factors | Age: Over 65 | ⇧ | | | |
| | Race: White | ⇧ | | | |
| | Hispanic/Latino | | | ⇧ | |
| | Educational attainment: Bachelor or higher | | | | ⇩ |
| | Household Income: <$50,000 | | ⇧ | ⇧ | |
| | Household Income: $100,000 to $199,999 | ⇧ | | | |
| Trip factors | Trip origin in NYC | | ⇧ | | |
| | Trip Destination in NYC | ⇩ | | | |
| | Medical device: Walking cane | ⇧ | ⇧ | | |
| | Medical device: Wheelchair | | | | ⇧ |
| | Log(Trip duration) | ⇩ | | | |
| | Log(Trip length) | ⇧ | | ⇧ | |
| | Trip purpose: Work | | ⇧ | | |
| | Trip purpose: Non-work | | | | ⇧ |
| | Trip category: Home-based | ⇩ | | | |
| | Trip category: Non-home-based | | ⇧ | | ⇧ |
| | Day of week: Weekday | ⇩ | | | |
| Environmental factors | Trip origin location: Rural | | | | ⇩ |
| | Trip destination location: Urban | | ⇧ | | |
| | Season: Summer | | | ⇩ | ⇧ |
| | Season: Fall | | ⇧ | | |
| | Season: Winter | | | ⇩ | |
| | Trip origin: Wind speed | | ⇧ | | |
| | Time of day: 7:00 am to 9:59 am | | | | ⇧ |

⇧ indicates increase and ⇩ indicates decrease in the probability of a mode choice.

5.2 Limitations and Future Work

The current study possesses several limitations. Firstly, it did not explicitly examine the transportation mode preferences of people with specific types of disabilities or residing in distinct geographic regions. Therefore, future research should undertake comparative analyses across various disability types and regions to identify commonalities and differences in the factors influencing transportation mode choice among this population.

Secondly, this study focused solely on TLD people who made at least one trip, thereby disregarding some factors that might compel people to remain at home and, consequently, hinder their access to essential needs. As a result, further research should investigate these constraints, encompassing both physical and mental barriers, and propose strategies to surmount them.

Thirdly, it is worth delving into several additional factors that warrant exploration. From the perspective of trip characteristics, the purpose of a trip, whether it involves grocery shopping, recreation, or errands, could hold distinct influences on mode preferences. Considering the supply side, the accessibility and service levels of paratransit options or accessible taxi services hold potential significance. For the demand side, the specific impairments a person faces, such as visual or hearing impairments, might also shape their mode choices. Furthermore, this study did not explore any interaction effects among the variables, such as whether income moderates the relationship between wheelchair use and public transport utilization. Thus, future studies should undertake more specific analyses to gain an in-depth understanding of these complex relationships, building upon the insights gleaned from this study.

Fourthly, our study did not examine the influence of built environment such as transit facilities on the mode preferences of TLD individuals. Nevertheless, it is important to recognize that factors within the built environment, such as the presence of sidewalks, pedestrian signals, and transit frequency, could hold considerable significance in shaping mode choices. Moreover, it is worth acknowledging that the impact of the built environment on mode preferences might vary across different regions, necessitating distinct modeling strategies to elucidate these variations (Ma et al., 2023). Subsequent research endeavors could delve deeper into unraveling the spatially heterogeneous effects of built environment factors on the mode preferences of TLD people.

Lastly, within NHTS, one survey question pertains to coping strategies. This aspect prompts further exploration into how socioeconomic indicators, weather conditions, and/or built environment attributes interplay to shape the coping strategies of TLD people. This could encompass investigating actions such as travel reduction, decreased bus utilization, or seeking companionship during travel. This aspect presents an avenue for further research, shedding light on the multifaceted dynamics underlying the travel behaviors of TLD people in response to various contextual cues.

## 6. Conclusions

This study identified mode choice influencing factors for people with travel-limiting disabilities (TLD). The modeling successfully identified influencing factors for each transportation mode alternative, including non-fixed effect variables that can vary among different groups of people with disabilities.

The study's found that low-income people with TLD are more likely to travel with public transport or walking, and potentailly more susceptible to the impact of weather conditions. In rural areas, transportation agencies could consider enhancing accessibility through the provision of paratransit services or other emerging technologies. It is also crucial for policymakers to give priority to accommodating infrastructure while designing transportation facilities. This is because people may be compelled to walk longer distances in the absence of suitable accommodation facilities for other modes of transportation.

This paper contributes to the existing literature by providing a comprehensive overview of the factors influencing mode choice for adults with TLD and highlighting the importance of accessibility and accommodation in transportation systems for people with disabilities. It also has important implications for transportation planners, policymakers, and disability advocates, as it can inform the development of more inclusive and accessible transportation systems. Understanding the factors influencing mode choice for people with disabilities can lead to more equitable transportation systems that meet the needs of all.

The results obtained from this study have a number of implications. First, affordability issues in relation to different modes of transportation for individuals with TLD need to be addressed, particularly those with lower incomes or people of color. This could be done through subsidies, discounts, or other incentives for individuals to use more affordable modes of transportation, such as public transport or walking. Second, improved accessibility to transportation facilities is required for individuals with TLD in rural areas. The improvement could be achieved through increased partnerships between rural agencies and transportation service providers. Additionally, the service could be customized based on the characteristics of residents within the area. Lastly, for individuals with different types of TLD, transportation mode access could be made more accommodating through better accessible boarding and onboarding facilities in public transit. Additionally, to ensure that extreme hot or cold weather conditions do not discourage individuals from walking, more covered walkways or indoor paths could be provided.

**Declaration of Generative AI and AI-assisted technologies in the writing process**

During the preparation of this work the authors used ChatGPT in order to refine certain language aspects. After using this tool/service, the authors reviewed and edited the content as needed and take full responsibility for the content of the publication.